\documentclass[withtimes]{easychair}

\usepackage{amsmath,amsfonts,comment}

\begin{document}
\title{Dataflow Graphs as Matrices\\ and Programming with Higher-order Matrix Elements}
\titlerunning{Dataflow Graphs as Matrices}
\author{Michael Bukatin\inst{1}\and Steve Matthews\inst{2}}

\institute{Nokia Corporation\\
Burlington, Massachusetts, USA\\ 
\email{bukatin@cs.brandeis.edu}
\and
Department of Computer Science\\
University of Warwick\\
Coventry, UK\\
\email{Steve.Matthews@warwick.ac.uk}}
\authorrunning{Bukatin and Matthews}

\maketitle

\begin{abstract}
We consider dataflow architecture for two classes of computations which admit taking linear combinations of execution runs:
probabilistic sampling and generalized animation. We improve the earlier technique of almost continuous program transformations
by adopting a discipline of bipartite graphs linking nodes obtained via general transformations and nodes obtained
via linear transformations which makes it possible to develop and evolve dataflow programs over these classes
of computations by continuous program transformations. The use of bipartite graphs allows us to
represent the dataflow programs from this class as matrices of real numbers and evolve and
modify programs by continuous change of these numbers. 

We develop a formalism for
higher-order dataflow programming for this class of dataflow graphs based on the higher-order matrix elements. Some of our software experiments
are briefly discussed.

\end{abstract}

\section{Introduction}

Because probabilistic sampling and generalized animation are both stream-based,
dataflow programming is a natural framework for this situation. Earlier we were able to
leverage the ability to take linear combinations of execution runs to obtain the
notion of {\em almost continuous transformation} of dataflow programs~\cite{MBukatinMatthews}.

There were three main sources of {\em benign discontinuities} in~\cite{MBukatinMatthews}:
addition of new subgraphs via an operation of {\em limited deep copy}, insertion of a new vertex and
an edge during the first stage of {\em S-insert}, and insertion of an edge during the second stage of
S-insert.

We now allow to decorate subgraphs with weights and require that new subgraphs created by
limited deep copy appear initially with zero weight subject to subsequent continuous evolution.

We also introduce the discipline of arranging and linking vertices as a bipartite graph, namely
that general transforms of fixed arity must point only at linear transforms of unlimited arity,
and vice versa.

This allows to eliminate the benign discontinuities mentioned above and obtain {\em continuous
program transformations}.

\subsection{Dataflow Graphs as Matrices}

It is convenient to have a situation where for any trajectory of program development or
evolution all parts of the program which might emerge preexist in a silent way.

The discipline of bipartite graphs actually makes this possible. We fix a particular
{\em signature} by taking a finite number of operations, each with its own fixed finite non-negative arity.

We take a countable number of copies of each {\em template operation} from the signature. Then we
have a countable set of inputs of those operations, $Y_j$, and a countable set of their
outputs, $X_i$.

Associate with each $Y_j$ a linear combination of all $X_i$ with real coefficients $a_{ij}$.
We require that no more than finite number of elements of the matrix $(a_{ij})$ are nonzero.

We often impose additional conditions, e.g. we often require that all $a_{ij}$ are
nonnegative, and that the sum of elements $a_{ij}$ associated with a particular matrix column $Y_j$
does not exceed 1.

Thus we have a countable-sized program, namely a countable dataflow graph, all but
a finite part of which is suppressed by zero coefficients. Any finite dataflow graph
over a particular signature can be embedded into a universal countable dataflow graph
over this signature in this fashion.

Hence we represent programs over a fixed signature as countable-sized real-valued matrices with
no more than finite number of nonzero elements, and any program evolution would be
a trajectory in this space of matrices.

\subsection{Multiple Types of Data Streams}

Originally this construction was envisioned
for the situation when each node $X_i$ and $Y_j$ has a data stream of the same type,
and all these streams are equipped with the same meaningful addition operation~\cite{MBukatinMatthews}.

The most important case here is when these are streams of real numbers which can be considered as
one-point generalized animations.

However, we might want to consider situations when there are multiple stream types associated with nodes.
Consider a situation when $Y_j$ and $X_i$ are of different types. If $a_{ij} = 0$, this is fine, because then
$a_{ij} \cdot X_i$ does not affect $Y_j$. If there is a default adapter $T_{ij}$ between these types, then other values of $a_{ij}$
are allowed, and $a_{ij} \cdot T_{ij}(X_i)$ is contributed to the sum. Otherwise the condition
$a_{ij} = 0$ has to be enforced ($a_{ij}$ are clamped at zero).

\subsection{Multiple Types of Addition}

One might want to also consider a situation where more than one meaningful
addition operation is possible within the same program (e.g. point-wise addition of generalized animations
and stochastic sum of the streams of probabilistic samples).

(Note that these additions are different from template operations: additions have unlimited arity and map infinite
tuples of $X$'s to $Y$'s, and template operations have finite fixed nonnegative arity and map 
finite tuples of $Y$'s to $X$'s.)

In the case of multiple types of addition, each input $Y_j$ of a template operation from the signature needs to be marked with the type of
the addition operation it uses. If necessary, similar template operations with different addition types should
be included in the signature separately.

One would typically include an identity transform for every type of data stream and for every type of addition
applicable to this particular type of data stream in order to provide a capability to group the sums
hierarchically.

\subsection{String-based Indices}\label{string_indices}

Rather than indexing our countable sets $\{X_i\}$ and $\{Y_j\}$ with numbers we are going to fix an alphabet and to index them with
strings from that alphabet. Of course, letters can be considered as digits in an appropriate base, and strings as the corresponding natural
numbers, but by indexing with strings we are trying to de-emphasize the order associated with this natural-numbers-based
interpretation and to encourage an implementation of sparse arrays based on dictionaries (hash tables), with the expectation
that zero elements of arrays and matrices would typically be omitted from their respective dictionaries.

In order to maintain our convention that the association between a particular $X_j$ and its template operation is fixed, and
that the association between $X_j$ and its corresponding arguments $Y_{k_1}, \dots, Y_{k_{n_j}} (n_j >= 0)$ is fixed and
that all flexibility is concentrated solely in the values $a_{ij}$ we adopt the following naming conventions.

We call a multiset (bag) of strings {\em prefix-free} if none of the strings in that multiset is a prefix of another string in that multiset.
(Note that a prefix-free multiset cannot have multiple occurrences of the same string.)
All template operations are given string names in such a way that the multiset consisting of all these names and the string $'arg'$
is prefix-free. A node $Y_j$ is indexed with a string $j$ which starts with the name of the template operation associated with
this node. We say that the string $j$ is the name of $Y_j$.
 A node $X_i$ which is one of the arguments for $Y_j$ is indexed with the string $i$ which is obtained by
concatenating the prefix $'arg1\ '$ ($'arg2\ '$ etc.) with the name of $Y_j$. We say that the string $i$ is the name of $X_i$.

\subsection{Higher-order Programming}

Because matrices, their columns, and their elements are vectors themselves, this is a platform with a variety of opportinuties
to create techniques for higher-order programming.

In this text we focus on one particular avenue for doing so, and this approach is centered around making
elements $a_{ij}$ higher-order (see Section~\ref{higher_elements}).

To quote from Section 1.1 of~\cite{MBukatinMatthewsLinear}
``A lot of expressive power of this architecture comes from the ability to have non-standard secondary structures on the set of points.
Points can be associated with vertices or edges of a graph, grammar rules, etc. One should be able to formulate
mechanisms of higher-order animation programming via variable illumination of elements of such structures."

What is being done in the present paper is an instance of this approach.

A more precise mathematical description of the way this kind of computational engine functions
is in Section~\ref{abstract_machine}.

Our first series of exercises in matrix-based dataflow programming which have been open-sourced simultaneously with the release of this
preprint is described in Section~\ref{simple_example}.

\section{Higher-order Matrix Elements}\label{higher_elements}

\subsection{Degrees of Order}

Here we classify varieties of $a_{ij}$.

\subsubsection{Not Properly Higher-order Elements}

A not properly higher-order element $a_{ij}$ is not associated with a stream of any particular node $X_k$.

\paragraph{Zero elements (zero-order).}

These coefficients are zero and are not typically included in the dictionaries implementing the matrix or its columns.

\paragraph{First-order elements.} Constant elements of the matrix $(a_{ij})$ are called first-order elements.

\paragraph{Variable elements (first-and-a-half order, sesquialteral-order elements).} Elements of the matrix $(a_{ij})$ which vary with
time belong to this class. Our exercises in higher-order dataflow have belonged to this class so far.

\subsubsection{Properly Higher-order Elements}

A properly higher-order element is associated with a datastream of a particular node $X_k$. The three types
above are still representable with higher-order elements. When a node is not turned on (and typically is
not in the appropriate dictionary), this is the case of zero-order element. When a node is constant value
with no arguments, this is the case of first-order matrix element. When a node is a variable stream, but does not
have argument nodes (so we are talking about a predefined/external variable stream), this is the case of sesquialteral-order element.

\paragraph{Specialized higher-order elements.}

When a node $X_k$ is computed by a template operation with at least one argument, this is the
case of a higher-order element. We call it specialized, because the operation which computes
this element is fixed, although it can be controlled via its argument(s). Evaluation of such an operation is immediately
effective, i.e. the newly computed coefficient is then immediately used to compute the
appropriate linear combination.

\paragraph{Fully higher-order elements.} In this case the node $X_k$ is computed by a template operation which
is the identity transform (one of the identity transforms, if multiple types of addition of streams of reals
are included in the signature). In this case, the value in question can be computed in a very flexible
manner from any linear combination of any transformations. However, the value is essentially
computed on the ``downswing" using the previous values of $(a_{ij})$, and becomes the new effective
value of the matrix only after the identity transform is applied to it. So if one needs to implement
changing the $a_{ij}$ via a particular specialized template operation, a one-cycle delay is
involved before the new value of $a_{ij}$ goes via a linear combination and via an identity transform and becomes effective.

\subsection{Embedding the Set of Matrix Coefficient Indices into the Set of Column Indices}\label{countable_embedding} 

An interesting question is whether one can have all matrix elements to be properly higher-order elements.
The answer is ``yes", and it is essentially based on the countability of the union of countable sets.
This countability allows us to take the set of indices of matrix coefficients, $(ij)$, and embed it into the set
of indices of matrix columns, $j$.

Here we are going to give an example of such an embedding which is sufficiently detailed to enable computer
implementation.

Without properly higher-order matrix elements, it was possible to avoid giving unique names for matrix elements.
For example, one could represent matrix columns as separate dictionaries and then just use the row names
for matrix elements, like we do in the example of Section~\ref{simple_example}.

With properly higher-order matrix elements one needs unique names for matrix elements. We start with
the naming scheme for columns and rows given in Section~\ref{string_indices} and given the name $N_j$ for
column $j$ and the name $M_i$ for row $i$, we define the name for the matrix element $(ij)$ as concatenation
$'('+N_j+')\#('+M_i+')'$. Now we should just reserve parentheses for the use within the names of matrix elements
only, but not within the ``originally present names", and this should be enough to avoid the name clashes.

\section{A More Precise Description of the Abstract Machine}\label{abstract_machine}

We only describe the machine for the situation where all nodes are streams of reals, and where one kind
of addition operation exists with the sum being component-wise addition. It is not difficult to generalize
appropriately.

Define an alphabet $\Sigma$ which does not contain parentheses and the number sign, $'()\#'$.

Denote the set of all finite strings over $\Sigma$ as $\Sigma^*$.

Define an operation as a tuple $F$ containing name (a string $N_F \in\Sigma^*$), arity (a non-negative integer $A_F$), and
a mathematical function $f_F : \mathbb{R}^{A_F} \rightarrow \mathbb{R}$ (if $A_F$ is zero, $f_F$ is simply a real number).

Define a signature as a finite set of operations, $S = \{F_1, \dots, F_n\}$. 

We require that the multiset $\{N_{F_1}, \dots, N_{F_n}, 'arg'\}$ is
prefix-free per Section~\ref{string_indices}.

We take $F_1$ to be the identity operation, $\langle 'id', 1, x \mapsto x \rangle$.

\subsection{Abstract Machine Without Properly Higher-order Matrix Elements}\label{abstract_basic}

The first case we consider is the case of sesquialteral-order matrix elements, which will be further
illustrated by the example in Section~\ref{simple_example}.

We introduce the partially defined map $Op$ from $\Sigma^*$ to $S$ (operations) and two partially
defined maps $X$ and $Y$ from $\Sigma^*$ to $\mathbb{N} \rightarrow \mathbb{R}$ (data streams).

We denote string concatenation with + and string representation of a non-negative integer $k$ as $str(k)$.

For $w \in \Sigma^*$ and for every $F_m \in S$ introduce 
$i= N_{F_m} + '\ '+w$. 
For all such $i$, $Op(i)$ and $X(i)$ (denoted as $X_i$) are defined and $Op(i) = F_m$.

For all such $i$, for all $k \in \{1, \dots, A_{F_m}\}$, introduce $j =\ 'arg' + str(k) + '\ ' + i$. 
For all such $j$, $Y(j)$ (denoted as $Y_j$) is defined. 

Now we have a dataflow graph representing the machine, and we need to define
the values of data streams $\mathbb{N} \rightarrow \mathbb{R}$ associated with
graph nodes $X_i$ and $Y_j$, which will define how the machine works.

Time starts with value 0 and increases by 1 at each machine step. For every $i$ and $j$
defined above there is a stream of values $\mathbb{N} \rightarrow \mathbb{R}$
associated with $a_{ij}$. Given that we are considering the case of
sesquialteral-order matrix elements, we assume that streams associated with
$a_{ij}$ are external to the program and are just generated by their
externally programmed generators.

At time 0 we start with all $X_i$ and $Y_j$ being the streams of one element,
which equals to 0 for all those streams. The values of $a_{ij}$ at the moment 0
are not important at this level of consideration, because we only give a mathematical
definition in this section and ignore the question of sparseness as an
implementation detail (of course, nothing would even fit a finite machine
without this implementation detail).

Then given streams $X_i$, $Y_j$, and $a_{ij}$ of length $t$, here is how
the components $t+1$ are defined (that is, computed by the abstract machine; so this is a definition by
induction with respect to time, and this definition describes how the machine works).

\paragraph{Step 1.} First, for all $i$, the value of $X_i$ at time $t+1$ is computed as follows.
One considers $Op(i)$, and for all  $k \in \{1, \dots, A_{Op(i)}\}$ and $j_k =\ 'arg' + str(k) + '\ ' + i$ one
takes the value $y_{j_k}$ as being the value of the stream $Y_{j_k}$ at moment $t$.
The value of $X_i$ at time $t+1$ then is computed (defined) as $f_{Op(i)}(y_{j_1} , \dots, y_{j_{A_{Op(i)}}})$.

\paragraph{Step 2.} Then the values of streams corresponding to $a_{ij}$ at the moment $t+1$ are externally generated.
The condition that only a finite number of those are different from 0 at the moment $t+1$ is observed.

\paragraph{Step 3.} Finally, for all $j$, the values of streams associated with $Y_j$ at the moment $t+1$ are computed as the linear
combinations $\Sigma_i a_{ij} \cdot X_i$, where $a_{ij}$ and $X_i$ are the values of the corresponding
data streams at the moment $t+1$.

\subsection{A More Abstract View}

Here is a more abstract way to describe what is going on in the previous subsection.

There are two (in general, countably dimensional) vector spaces $X$ and $Y$. There is
a fixed non-linear function $F : Y \rightarrow X$ induced by the signature. There is
a family of linear transformations $L_t: X \rightarrow Y$ parametrized with time
(this corresponds to sesquialteral-order of $L$).

We define streams of vectors $x_t$, $y_t$ from $X, Y$ parametrized with time by induction over time.
We define $x_0$ and $y_0$ to be zero vectors and $L_0$ is irrelevant.

At time $t+1$, first we take $x_{t+1} = F(y(t))$, then we note the current
$L_{t+1}$ and take $y_{t+1} = L_{t+1} (x_{t+1})$.

In the subsection which follows we use the embedding from Section~\ref{countable_embedding}
to incorporate $L$ into $X$.

Then at time $t+1$ we take $x_{t+1} = F(y(t))$ and among other things this yields $L_{t+1}$.
Then we take $y_{t+1} = L_{t+1} (x_{t+1})$. 

We occasionally parametrize $F$ as $F_t$ as well, to allow the dynamic tuning of the system
and to account for the stochastic factors in randomized template operations.

\subsection{Abstract Machine with all Fully Higher-order Matrix Elements}\label{fully_higher_order}

Here we modify the construction of Section~\ref{abstract_basic} in order to deliver what
was promised at the end of the previous subsection.We start with sets of names $I= \{i\}$ and
$J=\{j\}$ defined in  Section~\ref{abstract_basic}.

Then we define $I_0 = I$ and $J_0 = J$, and organize the induction as follows.

Given $I_n$ and $J_n$ we first use the scheme from Section~\ref{countable_embedding} to define
set of names of matrix elements: $A_{n+1} = \{'('+j+')\#('+i+')' | i \in I_n, j \in J_n\}$.

Then we use the fact that our signature has the identity operation and define
$I_{n+1} = I \bigcup \{'id\ '+ a | a \in A_{n+1}\}$ and $J_{n+1} = J \bigcup \{'arg1\ id\ '+ a | a \in A_{n+1}\}$.

Now consider the limit of this process, $I_{\infty} = \bigcup I_n$ and $J_{\infty} = \bigcup J_n$.

Then this limits is the fixed point of this process, namely that if we define\linebreak
$A_{\infty} = \{'('+j+')\#('+i+')' | i \in I_{\infty}, j \in J_{\infty}\}$, then
$I_{\infty} = I \bigcup \{'id\ '+ a | a \in A_{\infty}\}$ and $J_{\infty} = J \bigcup \{'arg1\ id\ '+ a | a \in A_{\infty}\}$.

Now the steps at the end of  Section~\ref{abstract_basic} are modified as follows.
Performing Step 1 yields values of streams corresponding to $a_{ij}$ at the moment $t+1$.
If only finite number of those values are different from 0 at the moment $t$, then only
a finite number of them would be different from 0 at the moment $t+1$ (for the detailed explanation
of the reasons for that see the last paragraph of the next subsection).

Then we can go ahead and perform Step 3.

\subsection{Sparseness Considerations}

The discussion of the abstract machines would not be complete without discussing how
to keep the amount of computations at each step finite (as, if possible, minimal).

For the setup of Section~\ref{abstract_basic}, the requirement is that if a matrix
element $a_{ij}$ is nonzero, then its linear combination needs to be evaluated and the
template operation computing $X_i$ and the template operation using $Y_j$ as an argument
both need to be evaluated, so the appropriate nodes should be included into the
dictionaries in question. When for a template operation with inputs $Y_1, \dots, Y_k$ and
output $X_i$ all matrix elements related to nodes $Y_1, \dots, Y_k$ and
output $X_i$ are zero, it is not recommended to evaluate such a template operation or
to include the nodes in question into dictionaries (although doing so is legal).

This means that when $a_{ij}$ becomes nonzero for the first time,
the $Y_j$ node and the template operation using this node as an input together with the other vertices
of this template operation must be added to the appropriate dictionaries, if it is not already there.
Similarly, the $X_i$ node and the template operation computing it together with the other vertices
of that template operation must be added to the appropriate dictionaries, if it is not already there,
and moreover, it needs to be retroactively computed (by evaluating the corresponding template operation
with zero values of its arguments) 
before the linear combination using  $a_{ij}$ can be evaluated.

The same considerations are valid for the setup of Section~\ref{fully_higher_order}. At any given moment
of time we only have a finite number of operations computing $a_{ij}$ being in the dictionaries and subject to being
evaluated. So only finite number of them might become nonzero for the first time at any given time moment.
Hence no more than a finite number of new nodes need to be added to the dictionaries on
each time step.

\section{A Simple Example}\label{simple_example}

This architecture is applicable to probabilistic sampling and to generalized animation.
We created an open source software prototype demonstrating the use
of those techniques for streams of real numbers implementing a family
of continuous cellular automata.

The software and associated videos have been released simultaneously with this
preprint as\linebreak {\tt aug\_20\_15\_experiment} in the Project Fluid repository~\cite{Fluid}.

Note that we cannot make these video previews quite faithful; 
video compression software introduces some distortion at the initial stage of uploading a video,
then it notices the problems with quality and offers to automatically fix lighting,
the overall result might be worse in some aspects and better in some other aspects
than the video natively produced by the software.

This is an example of a situation when a matrix-based dataflow program is used
in combination with other software. In this particular case, the video output
from the dataflow program is being passed through an external
lighting-enhancing software.

The set of template operations for this series of experiments is very simple: zero-arity
constants, {\tt black} and {\tt white} (-1 and 1), and a unary randomized propagator,
copying the value of its input $Y_j$ to its output $X_i$ with probability $p$ (a typical
value of which in this series of experiments is 0.995) and setting the output to zero
otherwise. The setup is essentially that of Section~\ref{abstract_basic}, except that
we allow several copies of a template operation to share an input when convenient
(mathematically this is equivalent to constraining several columns of a matrix to be the same).

In this series of experiments we use the dynamic nature of matrix elements very
sparingly, first having the weights concentrated on the output of the constants
(initialization phase) and then switching it (continuously, or, as in this series
of experiments, abruptly) to the output of the randomized propagators
forming a structure determining the connectivity
of the continuous cellular automaton in question. This architecture also allows to
continuously morph between a variety of connectivity structures, but
we have left this as a simple exercise to the reader.

We are observing a variety of emerging Turing structures with interesting dynamics
in this series of experiments, depending mostly on the connectivity pattern,
and also on the initialization configuration. In the spirit of non-equilibrium
thermodynamics, even when one starts with a uniform
initial pattern (all black or all white), the structures first emerge driven by the randomized propagators
and by the connectivity pattern, and then fade out to gray (zero value, the thermodynamic
equilibrium). If one includes the amplification during the visualization stage
(scaling the brightness of the image so that the maximal absolute value of a rendered point reaches 1),
then one can observe that the meaningful fine structures of the dynamic patterns
tend to persist indefinitely (in some cases the main motive is preserved, in
other cases the pattern is shifting and evolving unpredictably).

See the Fluid repository for further details.

\subsection{An Example with Properly Higher-order Matrix Elements}

The {\tt aug\_24\_15\_experiment} implements the same example using fully higher-order elements
(all but two matrix elements in that experiment are fully higher-order, while two of its elements are not properly
higher-order, and their change is controlled from the outside).

This experiment also features pause/resume facility controlled by the mouse key and the
space bar, improved brightness amplification, and an adjustment mechanism to stabilize the values
and prevent their relaxation to zero.

We recommend that the reader uses this experiment to further explore the system.

\section{Conclusion}

This architecture allows us to evolve dataflow programs in continuous fashion while those evolving programs
are running. This makes it possible to sample continuous trajectories in the space of
dataflow programs, in addition to the usual practices of sampling the syntax trees of programs.

The representation of programs as matrices of real numbers makes the task of program
learning more similar to the task of machine learning for more narrow and conventional
classes of models.

One way to introduce some higher-order mechanisms in dataflow programming (and in neural
networks) is via letting any particular data stream to propagate along multiple directions in the
graph, and to dynamically control the actual directions of propagation by setting the
multipliers along some of those paths to zero.

A somewhat unexpected result of the present preprint is that this particular higher-order
mechanism seems to be universal: all higher-order programming for a given class
of dataflow graphs over datastreams with linear combinations of execution runs built on top of a fixed
signature of template operations seems to be expressible via setting multipliers
along the paths of data propagation (in particular, setting all but the finite number of them
to zero).

{\bf Acknowledgments.} We would like to thank Ralph Kopperman, Lena Nekludova, and Josh Tenenbaum for helpful discussions and
to acknowledge collaboration with Lena Nekludova on software experiments.

\label{sect:bib}
\bibliographystyle{plain}
\bibliography{LinearModels}

\end{document}